\documentclass[12pt]{iopart}
\usepackage{epsfig}
\usepackage{citesort}

\newcommand{\gtrsim}{\,\rlap{\lower3.7pt\hbox{$\mathchar\sim$}}
\raise1pt\hbox{$>$}\,}
\newcommand{\lesssim}{\,\rlap{\lower3.7pt\hbox{$\mathchar\sim$}}
\raise1pt\hbox{$<$}\,}

\begin{document}

\title{The neutrino mass bound from WMAP-3, the baryon acoustic peak,
the SNLS supernovae and
the Lyman-$\alpha$ forest}
\author{Ariel Goobar$^1$, Steen Hannestad$^2$, Edvard M{\"o}rtsell$^3$, Huitzu Tu$^2$}
\address{$^1$ Department of Physics, Stockholm University, SE-106 91 Stockholm, Sweden}
\address{$^2$ Department of Physics and Astronomy, University of
Aarhus,\\ Ny Munkegade, DK-8000 Aarhus C, Denmark}
\address{$^3$ Department of Astronomy, Stockholm University, SE-106 91 Stockholm, Sweden}

\date{{\today}}

\begin{abstract}
We have studied bounds on the neutrino mass using new data from
the WMAP 3 year data, the Sloan Digital Sky Survey measurement of
the baryon acoustic peak, the Type Ia supernovae from SNLS, and
the Lyman-$\alpha$ forest. We find that even in the most general
models with a running spectral index where the number of neutrinos
and the dark energy equation of state are allowed to vary, the
95\% C.L. bound on the sum of neutrino masses is $\sum m_\nu \leq
0.62$ eV (95\% C.L.), a bound which we believe to be robust. In
the more often used constrained analysis with $N_\nu =3$, $w =
-1$, and $\alpha_s = 0$, we find a bound of 0.48 eV without using
the Lyman-$\alpha$ data. If Lyman-$\alpha$ data is used, the bound
shrinks to $\sum m_\nu \leq 0.2-0.4$ eV (95\% C.L.), depending
strongly on the Lyman-$\alpha$ analysis used.
\end{abstract}
\maketitle

\section{Introduction} 

In the past few years a new standard model of cosmology has been
established in which most of the energy density of the Universe is
made up of a component with negative pressure, generically
referred to as dark energy. The simplest form of dark energy is
the cosmological constant, $\Lambda$, which obeys $P_\Lambda =
-\rho_{\Lambda}$. This model provides an amazingly good fit to all
observational data with relatively few free parameters and has
allowed for stringent constraints on the basic cosmological
parameters.

The precision of the data is now at a level where observations of the
cosmic microwave background (CMB), the large scale structure (LSS) of
galaxies, and Type Ia supernovae (SNIa) can be used to probe
important aspects of particle physics such as neutrino properties.
Conversely, cosmology is now also at a level where unknowns from the
particle physics side can significantly bias estimates of cosmological
parameters.

The combination of all currently available data from neutrino
oscillation experiments suggests two important mass differences in
the neutrino mass hierarchy. The solar mass difference of $\Delta
m_{12}^2 \simeq 8 \times 10^{-5}$ eV$^2$ and the atmospheric mass
difference $\Delta m_{23}^2 \simeq 2.2 \times 10^{-3}$ eV$^2$
\cite{Maltoni:2004ei}. In the simplest case where neutrino masses
are hierarchical these results suggest that $m_1 \sim 0$, $m_2
\sim \Delta m_{\rm solar}$, and $m_3 \sim \Delta m_{\rm
atmospheric}$. If the hierarchy is inverted one instead finds $m_3
\sim 0$, $m_2 \sim \Delta m_{\rm atmospheric}$, and $m_1 \sim
\Delta m_{\rm atmospheric}$. However, it is also possible that
neutrino masses are degenerate, $m_1 \sim m_2 \sim m_3 \gg \Delta
m_{\rm atmospheric}$. Since oscillation probabilities depend only
on squared mass differences, $\Delta m^2$, such experiments have
no sensitivity to the absolute value of neutrino masses, and if
the masses are degenerate, oscillation experiments are not useful
for determining the absolute mass scale.

Instead, it is better to rely on kinematical probes of the neutrino
mass. Using observations of the CMB and the LSS of galaxies it has
been possible to constrain masses of standard model neutrinos. The
bound can be derived because massive neutrinos contribute to the
cosmological matter density, but they become non-relativistic so late
that any perturbation in neutrinos up to scales around the causal
horizon at matter-radiation equality is erased, i.e.\ the kinematics
of the neutrino mass influences the growth of structure in the
Universe. Quantitatively, neutrino free streaming leads to a
suppression of fluctuations on small scales relative to large by
roughly $\Delta P/P \sim - 8 \Omega_\nu/\Omega_m$
\cite{Hu:1997mj}. The density in neutrinos is related to the
number of massive neutrinos and the neutrino mass by
\begin{equation}
\Omega_\nu h^2 = \frac{\sum m_\nu}{93.2 \, {\rm eV}} = \frac{N_\nu
m_\nu}{93.2 \, {\rm eV}},
\end{equation}
where $h$ is the Hubble parameter in units of $100~{\rm km}~{\rm
s}^{-1}~{\rm Mpc}^{-1}$ and all neutrinos are assumed to have the
same mass. Such an effect would be clearly visible in LSS
measurements, provided that the neutrino mass is sufficiently
large, and a likelihood analysis based on the standard
$\Lambda$CDM model with neutrino mass as an added parameter in
general provides a bound for the sum of neutrino masses of roughly
$\sum m_\nu \lesssim 0.5-1$ eV, depending on exactly which data is
used \cite{Hannestad:2003xv,Elgaroy:2003yh,Allen:2003pt,Barger:2003vs,%
Hannestad:2003ye,Crotty:2004gm,Hannestad:2004nb,seljak,Fogli:2004as,%
Hannestad:2004bu,tegmark05}.

This should be compared to the present laboratory bound from $^3$H
beta decay found in the Mainz experiment, $m_{\nu_e} = \left(
\sum_i  |U_{ei}|^2 m_i^2\right)^{1/2} \leq 2.3$ eV \cite{mainz}.
It should also be contrasted to the claimed signal for
neutrinoless double beta decay in the Heidelberg-Moscow experiment
\cite{Klapdor-Kleingrothaus:2001ke,Klapdor-Kleingrothaus:2004wj,%
VolkerKlapdor-Kleingrothaus:2005qv}, which would indicate a value
of 0.1-0.9 eV for the relevant combination of mass eigenstates,
$m_{ee} = \left| \sum_j U^2_{ej} m_{\nu_j} \right|$. Some papers
claim that the cosmological neutrino mass bound is already
incompatible with this measurement.

However, those claims are based on a relatively limited parameter
space. Using a much more complicated model with a non-power law
primordial power spectrum it is possible to accomodate large
neutrino masses, provided that SNIa data is discarded
\cite{subir}. It was recently shown that there is a strong
degeneracy between neutrino mass and the equation of state of the
dark energy component when CMB, LSS, and SNIa data is considered.
The reason is that when the neutrino mass is increased, the matter
density must be increased accordingly in order not to conflict
with LSS data. This is normally excluded when the dark energy is
in the form of a cosmological constant. However, when the equation
of state parameter of the dark energy fluid, $w$, is taken to be a
free parameter, an increase in the matter density can be
compensated by a decrease in $w$ to more negative values.

Here we study how the degeneracy can be broken by the addition of
information provided by the measurement of the baryon acoustic beak in
the Sloan Digital Sky Survey data and the distances
to Type Ia supernovae provided by the SNLS data. We also study constraints derived
from combining all available cosmological data, including the
measurement of the small scale matter power spectrum from the
Lyman-$\alpha$ forest.

In the next section we discuss the cosmological data used, in
section 3 numerical results from the likelihood analysis are
provided, and finally section 4 contains a discussion.

\section{Cosmological data}

In order to probe the neutrino mass we have used some of the most
recent precision CMB, LSS and SNIa data.

\subsection{Type Ia supernovae.}
Observations of SNIa over a wide redshift range to determine
cosmological distances is perhaps the most direct way to probe the
energy content of the universe \cite{goobar95} and
have lead to a major paradigm shift
in cosmology
\cite{schmidt98,riess98,perlmutter99,knop03,tonry03,riess04}.  While
extremely succesful at showing the need for {\em dark energy} to
explain the derived distances, these data sets are plagued by
systematic uncertainties (e.g. dust extinction corrections,
K-corrections, calibration uncertainties, non-Ia contamination,
Malmquist bias, weak lensing uncertainties, etc.)
rendering them non-optimal for precision tests on
$w$. However, the situation has improved since the start of a
dedicated experiment, the SuperNova Legacy Survey (SNLS) at CFHT.
Thanks to the multi-band, rolling search technique, extensive
spectroscopic follow-up at the largest ground based telescopes
and careful calibration, this data-set is
arguably the best high-$z$ SNIa compilation to date, indicated by the
very tight scatter around the best fit in the Hubble diagram and the
careful estimate of systematic uncertainties by Astier et al., (2006)
\cite{astier06}.
The first year data of the SNLS collaboration includes 71 high redshift SNIa
in the redshift range $z=[0.2,1]$ and 44 low redshift SNIa compiled
from the literature but analysed in the same manner as the high-$z$
sample. We thus make only use of this new SNIa data set to minimize
the effects from systematic uncertainties in our analysis.

It should be noted that combining the results in Ref.~\cite{astier06}
with the Gold sample of Riess et al. (2004) \cite{riess04} as done in,
e.g. Refs.~\cite{Spergel:2006hy} and \cite{seljak2006} is not
straight forward.  First, the two data-sets use the same low-redshift
samples to anchor the Hubble diagram and are thus not
independent. Second, there are systematic differences in the way the
SNe are analysed, e.g. with respect to reddening corrections. In a
combined analysis, this would have to be adressed.

\subsection{Large Scale Structure (LSS).}

Any large scale structure survey measures the correlation function
between galaxies. In the linear regime where fluctuations are Gaussian
the fluctuations can be described by the galaxy-galaxy power spectrum
alone, $P(k) = |\delta_{k,gg}|^2$. In general, the galaxy-galaxy power
spectrum is related to the matter power spectrum via a bias parameter,
$b^2 \equiv P_{gg}/P_m$. In the linear regime, the bias parameter is
approximately constant, so up to a normalization constant, $P_{gg}$
does measure the matter power spectrum.

At present there are two large galaxy surveys of comparable size,
the Sloan Digital Sky Survey (SDSS)
\cite{Tegmark:2003uf,Tegmark:2003ud} and the 2dFGRS (2~degree
Field Galaxy Redshift Survey) \cite{2dFGRS}. Once the SDSS is
completed in 2006 it will be significantly larger and more accurate
than the 2dFGRS. In the present analysis we use data from both
surveys. In the data analysis we use only data points on scales larger
than $k = 0.15 h$/Mpc in order to avoid problems with non-linearity.

\subsection{Baryon acoustic oscillations (BAO).}

The acoustic oscillations at the time of CMB decoupling should be
imprinted also on the low-redshift clustering of matter and manifest
themselves as a single peak in the galaxy correlation function at
$\sim 100h^{-1}$ Mpc separation. Because of the large scale and small
amplitude of the peak, surveys of very large volumes are necessary in
order to detect the effect.

A power spectrum analysis of the final 2dFGRS data shows deviations
from a smooth curve at the scales expected for the acoustic
oscillations. The signature is much smaller than the corresponding
acoustic oscillations in the CMB but can be used to reject the case of
no baryons at 99\% C.L. \cite{Cole2005}.

The SDSS luminous red galaxy (LRG) sample contains 46,748 galaxies
with spectroscopic redshifts $0.16<z<0.47$ over 3816 square
degrees. Even though the number of galaxies is less than the 2dFGRS or
the main SDSS samples, the large survey volume ($0.72h^{-3}$ Gpc$^3$)
makes the LRG sample better suited for the study of structure on the
largest scales. The LRG correlation function shows a significant bump
at the expected scale of $\sim 150$ Mpc which, combined with the
detection of acoustic oscillations in the 2dFGRS power spectrum,
confirms our picture of LSS formation between the epoch of CMB
decoupling and the present.

For a given cosmology, we can predict the correlation function (up to
an amplitude factor which is marginalized over) and compare with the
observed LRG data. The observed position of the peak will depend on
the physical scale of the clustering and the distance relation used in
converting the observed angular positions and redshifts to positions
in physical space. The characteristic physical scale of the acoustic
oscillations is given by the sound horizon at the time of CMB
decoupling and depends most strongly on the combination $\Omega_m
h^2$. The conversion between positions in angular and redshift space
to positions in physical space will cause the observed correlation
scale to depend on the distance combination
\begin{equation}\label{eq:D}
D_V(z) = \left[ D_M(z)^2 {cz\over H(z)}\right]^{1/3} \, ,
\end{equation}
where $H(z)$ is the Hubble parameter and $D_M(z)$ is the comoving
angular diameter distance.

Our approach to fit the data has been to calculate the matter
power spectrum for a given model, then Fourier transforming it to
obtain the 2-point correlation function, $\xi(r)$. This
correlation function has been fitted to the SDSS data using the
full covariance matrix given by \cite{Eisenstein2005}.

In terms of the simple parametrization provided by
\cite{Eisenstein2005} in terms of the parameter
\begin{equation}
A \equiv D_V(z) {\sqrt{\Omega_m H_0^2}\over z c},
\end{equation}
we find that the SDSS constraint can approximately be written as
\begin{equation}
    A=0.469\left(\frac{n}{0.98}\right)^{-0.35} (1+0.94 f_\nu)\pm 0.017 \, ,
\end{equation}
where $f_\nu = \Omega_\nu/\Omega_m$.

\subsection{The Lyman-$\alpha$ forest.}

Measurements of the flux power spectrum of the Lyman-$\alpha$
forest has been used to measure the matter power spectrum on small
scales at large redshift. By far the largest sample of spectra
comes from the SDSS survey. In Ref.~\cite{mcdonald} this data was
carefully analyzed and used to constrain the linear matter power
spectrum. The derived amplitude is $\Delta^2 (k=0.009 \, {\rm km
\, s}^{-1}, z=3) = 0.452^{+0.07}_{-0.06}$ and the effective
spectral index is $n_{\rm eff} = -2.321^{+0.06}_{-0.05}$. The result
has been derived using a very elaborate model for the local
intergalactic medium, including full N-body simulations. It has been
shown that using the Lyman-$\alpha$ data does strengthen the bound on
neutrino mass significantly. However the question remains as to the
level of systematic uncertainty in the result. The same data has been
reanalyzed by Seljak et al. \cite{seljak2006} and Viel et
al. \cite{Viel:2005eg,Viel:2005ha,viel2006}, with somewhat different
results. The normalization found in
Refs.~\cite{Viel:2005eg,Viel:2005ha,viel2006} is lower than that found
by Ref.~\cite{mcdonald} which has significant influence on the
inferred neutrino mass bound.

Our Lyman-$\alpha$ analysis uses only the effective parameters
$\Delta^2(k)$ and $n_s$. The resulting bound on $m_\nu$ using these
parameters was shown in Ref.~\cite{Fogli:2004as} to be very similar to
those obtained in a full analysis. Furthermore we stress that using
these parameters makes it very simple to test the influence on
different assumptions about the Lyman-$\alpha$ data on the neutrino
mass bound.

\subsection{Cosmic Microwave Background (CMB).}

The temperature fluctuations are conveniently described in terms
of the spherical harmonics power spectrum $C_{T,l} \equiv \langle
|a_{lm}|^2 \rangle$, where $\frac{\Delta T}{T} (\theta,\phi) =
\sum_{lm} a_{lm}Y_{lm}(\theta,\phi)$.  Since Thomson scattering
polarizes light, there are also power spectra coming from the
polarization. The polarization can be divided into a curl-free
$(E)$ and a curl $(B)$ component, yielding four independent power
spectra: $C_{T,l}$, $C_{E,l}$, $C_{B,l}$, and the $T$-$E$
cross-correlation $C_{TE,l}$.

The WMAP experiment has reported data on $C_{T,l}$ and ,
$C_{EE,l}$, and $C_{TE,l}$ as described in
Refs.~\cite{Spergel:2006hy,Hinshaw:2006ia,Page:2006hz} . We have
performed our likelihood analysis using the prescription given by
the WMAP
collaboration~\cite{Spergel:2006hy,Hinshaw:2006ia,Page:2006hz}.

We furthermore use the newly published results from the Boomerang
experiment \cite{Jones:2005yb,Piacentini:2005yq,Montroy:2005yx}
which has measured significantly smaller scales than WMAP.

\section{Likelihood analysis}

Using the presently available precision data we have performed a
likelihood analysis for the neutrino mass.

As our framework we choose a flat dark energy model with the
following free parameters: $\Omega_m$, the matter density, the
curvature parameter, $\Omega_b$, the baryon density, $w$, the dark
energy equation of state, $H_0$, the Hubble parameter, $n_s$, the
spectral index of the initial power spectrum, $\alpha_s$, the
running of the primordial spectral index, $N_\nu$, the effective
number of neutrino species, and $\tau$, the optical depth to
reionization. Finally, the normalization, $Q$, of the CMB data,
and the bias parameter $b$ are used as free parameters. The dark
energy density is given by the flatness condition $\Omega_{\rm DE}
= 1 - \Omega_m - \Omega_\nu$. Including the neutrino mass our
benchmark model has 11 free parameters. We also test more
restricted parameter spaces in order to probe parameter
degeneracies.

The priors we use are given in Table~\ref{table:priors}. The prior
on the Hubble constant comes from the HST Hubble key project value
of $h_0 = 0.72 \pm 0.08$ \cite{freedman}, where $h_0 = H_0/100 \,
{\rm km} \, {\rm s}^{-1} \, {\rm Mpc}^{-1}$.

\begin{table}
\begin{center}
\begin{tabular}{lcl}
\hline \hline parameter & prior\cr
\hline $\Omega=\Omega_m+\Omega_{\rm DE} + \Omega_\nu$&1&Fixed\cr
$\Omega_m$ & 0--1 & Top hat \cr
 $h$ & $0.72 \pm 0.08$&Gaussian
\cite{freedman}\cr $\Omega_b h^2$ & 0.014--0.040&Top hat\cr
$N_\nu$ & 0 -- 10 & Top hat \cr $w_{\rm DE}$ & -2.5 -- -0.5 & Top
hat \cr $n_s$ & 0.6--1.4& Top hat\cr $\alpha_s$ & -0.5 -- 0.5 &
Top Hat \cr $\tau$ & 0--1 &Top hat\cr $Q$ &
--- &Free\cr $b$ & --- &Free\cr
$\sum m_\nu$ & ---  & Fitted over \cr \hline \hline
\end{tabular}
\end{center}
\caption{The different priors on parameters used in the likelihood
analysis.} \label{table:priors}
\end{table}

When calculating constraints, the likelihood function is found by
minimizing $\chi^2$ over all parameters not appearing in the fit
(i.e.\ over all other parameters than $m_\nu$).

\subsection{Results}

\begin{figure}
\hspace*{1.5cm}\includegraphics[width=100mm]{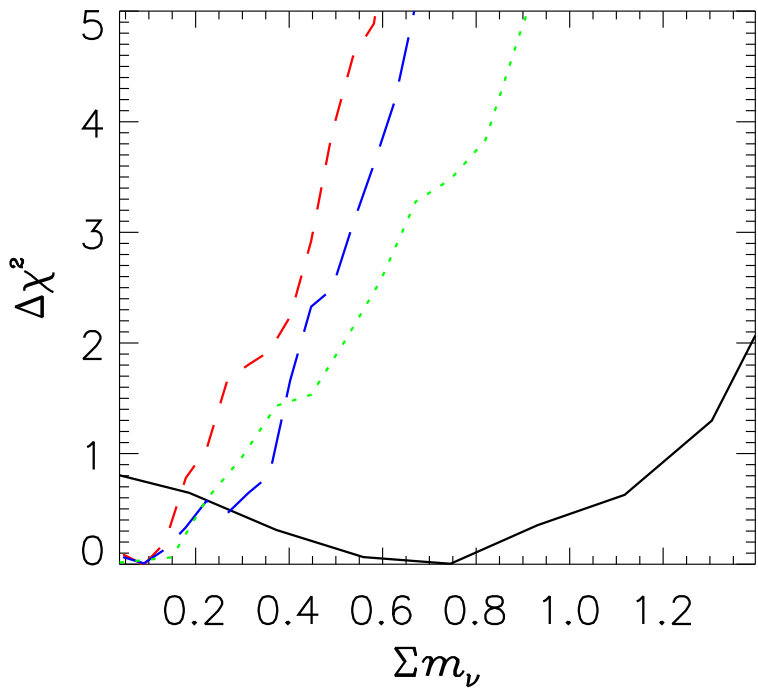}
\caption{The value of $\Delta \chi^2$ as a function of $\sum
m_\nu$ for various different data sets used for the full
11-dimensional parameter space. The curves are identical to the
cases described in Table 2: The full curve is case 1, the
long-dashed is case 2, the dotted is case 3, and the dashed is
case 4.} \label{fig:fig1}
\end{figure}

In Fig.~\ref{fig:fig1} we show the one dimensional likelihood
function for the neutrino mass for various different data sets,
using the full 11-dimensional parameter space.

By far the most conservative is for the case which includes only
CMB, LSS, and SN-Ia data. Because the parameter space is larger
than the one used in Ref.~\cite{Hannestad:2005gj}, the constraint
of 1.72 eV is, however, stronger because the addition of new
WMAP-3 data breaks some of the degeneracy \footnote{Note, however,
that if only WMAP-3 data is used, the inferred bound of 2 eV
\protect\cite{fukugita} is almost identical to what is found from
WMAP-1. This means that if used alone WMAP-3 does not add
significant new information}. It should be noted though that the
bound is weaker than the one shown in Fig.~18 of
Ref.~\cite{Spergel:2006hy}. The main reason for this is most
likely that $N_\nu$ and $\alpha_s$ is allowed to vary.

When the BAO data is added it has the effect of essentially
breaking the degeneracy between $m_\nu$ and $w$. The reason is
that the BAO measurement is almost orthogonal to the SNIa
measurement in the $[\Omega_m,w]$-plane. For this case the bound
shrinks to 0.62 eV, a factor of three improvement.

We also test the inclusion of SDSS Lyman-$\alpha$ data in the fit.
When BAO data is not included the Lyman-$\alpha$ data itself does
break some of the degeneracy, to a level where the formal bound is
0.83 eV. Including both BAO and Lyman-$\alpha$ data does lead to a
significant improvement, the combined bound being 0.49 eV at
2$\sigma$.

It is interesting to see how this bound compares with the value
obtained in the standard $\Lambda$CDM+$m_\nu$ parameter space
(which has 8 parameters). Including the BAO data we find an upper
bound of 0.48 eV, and with the SDSS Lyman-$\alpha$ data we find
0.35 eV. Combining all data leads to a bound of 0.27 eV. It should
be noted that this bound is somewhat weaker than what is found in
Ref.~\cite{seljak2006}, presumably because of the new treatment of
Lyman-$\alpha$ in Ref.~\cite{seljak2006}.

\begin{table}
\begin{center}
\begin{tabular}{lc}
\hline \hline Data & $m_\nu$ (95\% C.L.) \cr
\hline 1: CMB, LSS, SNIa & 1.72 eV \cr 2: CMB, LSS, SNIa, BAO &
0.62 eV \cr 3: CMB, LSS, SNIa, Ly-$\alpha$ & 0.83 eV \cr 4: CMB,
LSS, SNIa, BAO, Ly-$\alpha$ & 0.49 eV  \cr \hline \hline
\end{tabular}
\end{center}
\caption{Best fit $\chi^2$ values for the four different analyses
presented in Fig.~\ref{fig:fig1}, in all cases based on the full
11-dimensional parameter space. } \label{table:fits}
\end{table}

\begin{table}
\begin{center}
\begin{tabular}{lc}
\hline \hline Data & $m_\nu$ (95\% C.L.) \cr
\hline 1: CMB, LSS, SNIa & 0.70 eV  \cr 2: CMB, LSS, SNIa, BAO &
0.48 eV  \cr 3: CMB, LSS, SNIa, Ly-$\alpha$ & 0.35 eV  \cr 4: CMB,
LSS, SNIa, BAO, Ly-$\alpha$ & 0.27 eV  \cr \hline \hline
\end{tabular}
\end{center}
\caption{Best fit $\chi^2$ values for the three different analyses
presented in Fig.~\ref{fig:fig2}, in all cases based on the
restricted 8-dimensional parameter space with $N_\nu=3$, $w=-1$,
and $\alpha_s = 0$. } \label{table:fits2}
\end{table}

The strength of the bound including Lyman-$\alpha$ therefore depends
crucially on the assumed uncertainty in the measurement of
$\Delta^2(k)$. We have not used the most recent analysis of the SDSS
data by Seljak et al. \cite{seljak2006}, but in order to compare
roughly with their result we have added Lyman-$\alpha$ data that
mimics what is shown in their Fig.~1. For this model we find a bound
of 0.20 eV, in good agreement with their result of 0.17 eV.

However, we also note that using the Lyman-$\alpha$ result from
Viel et al. \cite{Viel:2005eg,Viel:2005ha,viel2006} which, when
combined with WMAP-3, has a best-fit normalization (note that this
is mainly due to the fact that the high resolution Lyman-$\alpha$
data used in Refs.~\cite{Viel:2005eg,Viel:2005ha,viel2006} has
larger error bars) about 2$\sigma$ lower than in
Ref.~\cite{seljak2006}, would lead to a different bound on $\sum
m_\nu$. To test this we have also run a likelihood analysis using
the Viel et al. data on $\Delta^2(k)$ and found a bound of 0.40
eV. This is the limit we inferred by using the derived
cosmological parameters from  the analysis of the SDSS data of
\cite{Viel:2005eg,Viel:2005ha,viel2006}, which consists of a
Taylor expansion of the flux power around a best fit model (SDSS-d
in their tables). With this analysis the bound from Lyman-$\alpha$
adds relatively little information to that obtained from BAO data.
We show the $\chi^2$ curves for this analysis in
Fig.~\ref{fig:fig3}.

This result leads to the inevitable conclusion that while the
Lyman-$\alpha$ data holds the potential to provide crucial
information on the neutrino mass, there seem to be unresolved
systematic issues related to the conversion of the measured flux
power spectrum to $\Delta^2(k)$. Any bound derived from the use of
the Lyman-$\alpha$ power spectrum should therefore be treated with
some caution. This is especially true since in the Seljak et al.
\cite{seljak2006} analysis there is a tension between the WMAP and
Lyman-$\alpha$ normalization on small scales of more than
2$\sigma$. While this could be statistical in nature, the
difference with respect to the Viel et al. \cite{viel2006}
analysis of the same data suggests a possible systematic origin.

\begin{figure}
\hspace*{1.5cm}\includegraphics[width=100mm]{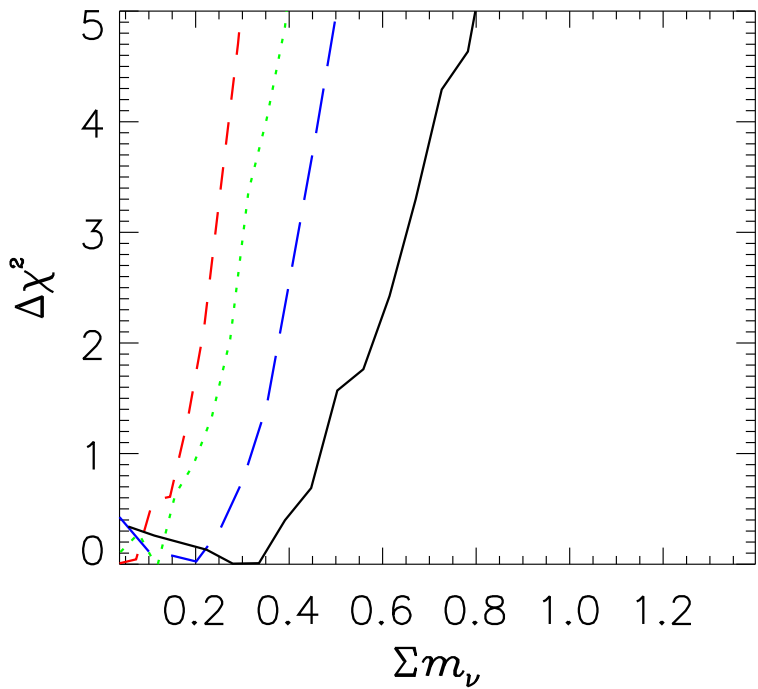}
\caption{The value of $\Delta \chi^2$ as a function of $\sum
m_\nu$ for various different data sets used for the restricted
8-dimensional parameter space with $N_\nu=3$, $w=-1$, and
$\alpha_s = 0$. The curves are identical to the cases described in
Table 3: The full curve is case 1, the long-dashed is case 2, the
dotted is case 3, and the dashed is case 4.} \label{fig:fig2}
\end{figure}

\begin{figure}
\hspace*{1.5cm}\includegraphics[width=100mm]{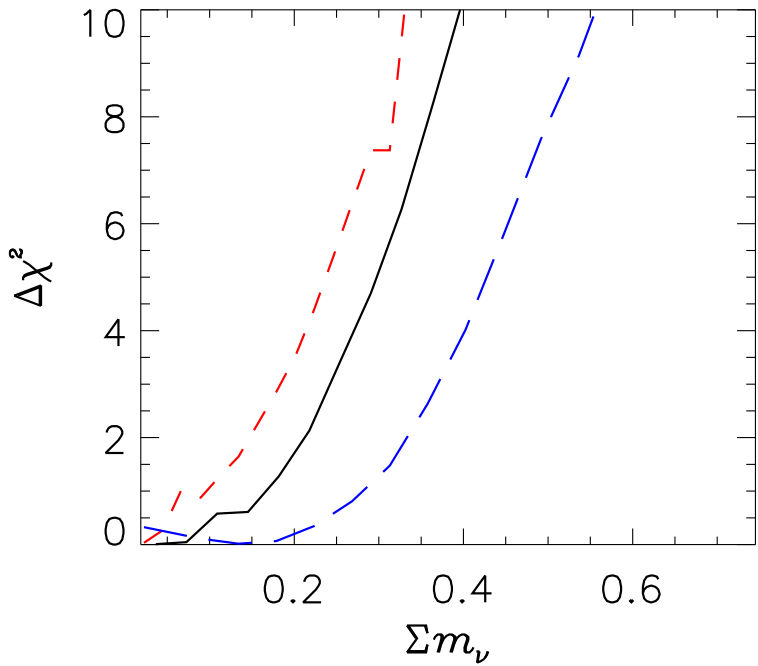}
\caption{The value of $\Delta \chi^2$ as a function of $\sum
m_\nu$ for the restricted 8-dimensional parameter space with
$N_\nu=3$, $w=-1$, and $\alpha_s = 0$ for different assumptions about
the Lyman-$\alpha$ data. The full line is for the Ref.~\cite{mcdonald}
data, the dashed is for the approximate analysis with the data in
Ref.~\cite{seljak2006}, and the long-dashed is for the data in
Ref.~\cite{viel2006}.} \label{fig:fig3}
\end{figure}

\section{Discussion}

We have calculated the bound on the sum of light neutrino masses from
a combination of the most recent cosmological data. If only CMB, LSS,
and SNIa data is used, we find a strong degeneracy between $\sum
m_\nu$, $N_\nu$, and $w$ which severely limits the ability of this
data to constrain the neutrino mass. However, once data from the SDSS
measurement of the baryon acoustic peak is included, this degeneracy
is broken because it measures $\Omega_m$ and $w$ very precisely. The
derived bound and best fit model is compatible with what can be
derived from observations of the Lyman-$\alpha$ forest, but is likely
much less affected by systematics. The bound is $\sum m_\nu \lesssim
0.6$ eV even for a very general 11 parameter cosmological model. If
data from the Lyman-$\alpha$ forest is added, the bound is $\sum m_\nu
< 0.2-0.4$ eV (95\% C.L.), depending on the specific Lyman-$\alpha$
analysis used.

Beyond SDSS, future galaxy redshift surveys will achieve a larger
effective volume and go to higher redshifts (see for instance
Ref.~\cite{Eisenstein:2003qy}). The prospects for them to constrain
$m_\nu$ have been studied in
e.g. Refs.~\cite{Lesgourgues:2004ps,Takada:2005si}. With the help of
the BAO measurement, the angular diameter distance and the Hubble
parameter $H (z)$ can be measured to percent level. This enables $w$
to be determined within $\sim
10\%$~\cite{Seo:2003pu,Matsubara:2004fr,Wang:2006qt}, breaking the
$m_\nu-w$ degeneracy.

Another powerful probe in the future is weak gravitational
lensing. It traces directly the mass distribution in a wide range
of scales, and thus does not suffer from the light-to-mass bias in
large scale structure surveys, while still being sensitive to
effects from neutrino early free-streaming. Cosmic shear
measurement with tomographic redshift binning of sources galaxies
is particularly effective in constraining dark energy parameters.
The degeneracy between $m_\nu$ and $w$ can then be broken in a
similar way as BAO does. Systematics herein arise from photometric
redshift uncertainties and shear calibration errors, which are
expected to be under control to the required accuracy. The major
uncertainty comes from estimating the nonlinear part of the matter
power spectrum~\cite{Smith:2002dz,Cooray:2002di}. It is found that
future ground- and space-based surveys such as CFTHLS
\cite{cfhtls}, SNAP \cite{snap} and LSST \cite{lsst}, combined
with future CMB measurement from the Planck Surveyor
\cite{planck}, can constrain neutrino mass to $\sigma (\sum m_\nu)
= 0.025 - 0.1$ eV~\cite{Song:2003gg,Hannestad:2006}.

CMB photons from the last scattering surface at $z \approx 1100$
are also deflected by the large scale structure at $z \lesssim 3$.
Extracting the weak lensing information encoded in the CMB signal
will significantly enhance the sensitivity of CMB experiments to
small neutrino mass~\cite{Kaplinghat:2003bh,Lesgourgues:2005yv}.
The forecast error $\sigma (\sum m_\nu)$ obtained in
Ref.~\cite{Lesgourgues:2005yv} is $\sim 0.15$ eV for Planck and
SAMPAN \cite{sampan}, and $0.035$ eV for the future Inflation
Probe project \cite{inflationprobe}. To summarize, we expect near
future cosmological observations to pin down the total neutrino
mass to a precision better than $0.1$ eV, the level of the
inverted hierarchy.

\section*{Acknowledgments} 

Use of the publicly available CMBFAST package~\cite{CMBFAST} and
of computing resources at DCSC (Danish Center for Scientific
Computing) are acknowledged. The authors would also like to thank
Daniel Eisenstein for helpful comments on BAO data and Matteo Viel
for many helpful discussions on the use of Lyman-$\alpha$ data.
A.G. would like to thank the G\"{o}ran Gustafsson Foundation for
financial support and E.M. thanks the Royal Swedish Academy of
Sciences for finacial support.

\vspace*{2cm}

\section*{References} 

\end{document}